# Unusual suppression of the superconducting energy gap and critical temperature in atomically thin NbSe$_2$

E. Khestanova[1,&], J. Birkbeck[1,&], M. Zhu[1,&], Y. Cao[1,2], G. L. Yu[1], D. Ghazaryan[1], J. Yin[1], H. Berger[3], L. Forró[3], T. Taniguchi[4], K. Watanabe[4], R. V. Gorbachev[1,2], A. Mishchenko[1], A. K. Geim[1,2], I. V. Grigorieva[1*]

[1]School of Physics & Astronomy, University of Manchester, Oxford Road, Manchester M13 9PL, United Kingdom
[2]National Graphene Institute, University of Manchester, Manchester M13 9PL, United Kingdom
[3]Laboratory of Physics of Complex Matter (LPMC), Ecole Polytechnique Fédérale de Lausanne, 1015 Lausanne, Switzerland
[4]National Institute for Materials Science, 1-1 Namiki, Tsukuba, 305-0044 Japan[5]

& These authors contributed equally

**It is well known that superconductivity in thin films is generally suppressed with decreasing thickness. This suppression is normally governed by either disorder-induced localization of Cooper pairs, weakening of Coulomb screening, or generation and unbinding of vortex-antivortex pairs as described by the Berezinskii-Kosterlitz-Thouless (BKT) theory. Defying general expectations, few-layer NbSe$_2$ – an archetypal example of ultrathin superconductors – has been found to remain superconducting down to monolayer thickness. Here we report measurements of both the superconducting energy gap $\Delta$ and critical temperature $T_C$ in high-quality monocrystals of few-layer NbSe$_2$, using planar-junction tunneling spectroscopy and lateral transport. We observe a fully developed gap that rapidly reduces for devices with the number of layers $N \leq 5$, as does their $T_C$. We show that the observed reduction cannot be explained by disorder, and the BKT mechanism is also excluded by measuring its transition temperature that for all $N$ remains very close to $T_C$. We attribute the observed behavior to changes in the electronic band structure predicted for mono- and bi- layer NbSe$_2$ combined with inevitable suppression of the Cooper pair density at the superconductor-vacuum interface. Our experimental results for $N > 2$ are in good agreement with the dependences of $\Delta$ and $T_C$ expected in the latter case while the effect of band-structure reconstruction is evidenced by a stronger suppression of $\Delta$ and the disappearance of its anisotropy for $N = 2$. The spatial scale involved in the surface suppression of the density of states is only a few angstroms but cannot be ignored for atomically thin superconductors.**

Superconductivity in two dimensions (2D) has been a subject of long-standing interest and debate, with the central question of whether superconductivity survives down to a monolayer thickness and what determines the superconducting (SC) transition temperature, $T_C$, in this case[1-12]. Until recently studies of ultrathin superconductors were limited to either amorphous thin films with strong disorder[5-9] or epitaxial layers[3,4,10]. In ultrathin amorphous films the SC pairing competes with electron localization and a transition to the insulating state is observed at the quantum resistance for Cooper pairs, $h/4e^2$ (ref.[6-9]). On the other hand, in epitaxial layers, such as Pb on Si(111) or FeSe on SrTiO$_3$ an essential role is played by interactions with the substrate[4,10]. Only recently has it become possible to study superconductivity in isolated mono- and few- layer crystals, thanks to advances in exfoliation and encapsulation techniques that allow studying ultrathin superconductors unstable in air while controlling their thickness with monolayer precision[12-15]. This showed that, whereas superconductivity survives down to a monolayer[13,16-18], a common trend for all ultrathin superconductors is a decreasing $T_C$ as the film thickness is reduced. In exfoliated transition metal dichalcogenides (TMDs), such as superconducting 2H-NbSe$_2$ studied here, $T_c$ decreases gradually down to the bilayer thickness[13,15,16] and then drops rather markedly for monolayers[13,16,19]. A similar behavior was also found in electrostatically doped 2H-MoS$_2$ (ref.[20]). Remarkably, for NbSe$_2$ with $N \geq 2$ ($N$ is the number of layers) the measured $T_c(N)$ are reproducible, with different groups reporting very similar values[13,15,16]. The only exception is monolayer NbSe$_2$ that exhibited a wider range of $T_c$ (refs.[13,16,18,19,21]) and was often found to be oxidized or contain a high density of defects'[22], indicating sensitivity of its $T_c$ to fabrication conditions. The high reproducibility of the measured $T_c(N)$ suggests a fundamental mechanism for its suppression, independent of the preparation conditions, as long as any exposure to oxidants during fabrication is avoided. To explain the universal suppression of $T_c$ in crystalline few-layer superconductors, several different mechanisms have been proposed[2,3,19,21], and further experimental evidence is needed to distinguish between those possibilities. Critically, very little information is available[23] on the corresponding evolution of the SC energy gap.

Here we use lateral transport and tunneling spectroscopy to study high-quality atomically thin 2H-NbSe$_2$ (further referred to simply as NbSe$_2$) with the number of layers from $N$=50 down to the monolayer. This allowed us to measure the dependence of the SC energy gap $\Delta$ on $N$ and to relate it to the observed $T_c$. All our NbSe$_2$ samples are high-quality monocrystals with relatively high residual resistance ratios (RRR = 2-7) and the mean-free path $l$ for charge carriers well in excess of the bulk coherence length $\xi_0 \approx 8$ nm (ref.[24]). Accordingly, electron scattering at low temperatures is dominated by surfaces rather than defects or impurities. For all $N$ we have found a fully developed SC gap that becomes gradually smaller in thinner crystals, following a very similar $N$ dependence to that of $T_c$. We attribute the observed decrease of $\Delta$ and $T_C$ mostly to the additional surface energy contribution imposed by the boundary condition on the electron wavefunction at the surfaces of ultrathin crystals. While negligible in bulk superconductors, this surface effect becomes dominant in atomically thin superconductors, as explained below. Another important contribution is changes in the electronic structure for few-layer superconductors which are witnessed in our experiments through a marked reduction in the anisotropy of $\Delta$.

We first characterized our few-layer NbSe$_2$ crystals using lateral transport and Hall measurements (the number of layers, $N$, refers to the unit cell of NbSe$_2$, i.e., 3 atomic layers). To this end, we have fabricated a number of devices where NbSe$_2$ crystals with different $N$ were exfoliated from bulk crystals in the inert atmosphere of a glove box and encapsulated with hexagonal boron nitride (hBN) or graphene. Here we used the same fabrication procedure as in ref.[13] where it was demonstrated that encapsulation prevents exposure of NbSe$_2$ to air and solvents during fabrication steps (such exposure is known to destroy NbSe$_2$ superconductivity[25,26]). Further details of device fabrication are given in Supplementary Note 1. We found that the device behavior (sheet resistance $R$, temperature dependence, etc.) was independent of the encapsulating material (hBN or graphene), presumably because of little charge transfer between graphene and NbSe$_2$ and a relatively high resistivity of the former. Importantly, for all devices we were able to use four-contact geometry (Supplementary Fig. 1e), which allowed us to eliminate the contact resistance and accurately measure $R$.

Figure 1a shows typical temperature dependences of the sheet resistance $R$ normalized by its value at 8 K, and Fig. 1b shows the low-$T$ parts of the same (not-normalized) curves corresponding to the SC transitions for different $N$. The $R(T)$ curves such as those in Fig. 1b were used to extract $T_c(N)$ (inset of Fig. 1b), where we employed the mean-field definition of $T_C$ corresponding to $R = R_n/2$ ($R_n$ is the low-$T$ normal-state resistance). Similar to the earlier reports[16,25,27], $T_c$ in our 2D NbSe$_2$ is gradually reduced with decreasing thickness, with the largest drop seen for the monolayer. The shallower slopes of $R(T)$ for $N$ =1-3 compared to larger $N$ (Fig. 1a) imply stronger electron scattering in thinner samples, as expected due to an increasing contribution from their surfaces[28,29]. Nevertheless, the normal-state sheet resistance remains low, namely, $R_n \approx 90$ Ω for the bilayer and $\approx 240$ Ω for the monolayer (see Fig. 1), much smaller than the resistance quantum. This rules out the suppression of $T_c$ by disorder, which was observed in amorphous films of comparable thickness but with much higher resistivity[7,8] (see below). The low $R_n$ also indicates that the crystals' bulk is unaffected by exfoliation and fabrication procedures, in agreement with monocrystallinity and the low defect density in encapsulated 2D NbSe$_2$, as observed by transmission electron and scanning tunneling microscopy.

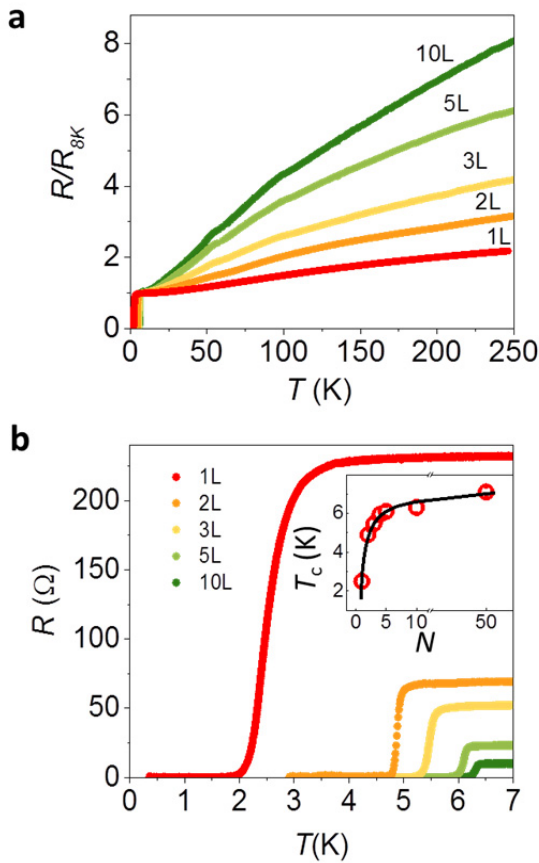

**Figure 1. Superconducting transition in few-layer NbSe$_2$ by transport measurements.** **(a)** Sheet resistance $R$ as a function of $T$ for different number of layers (shown as labels) normalised to the low-temperature resistance at 8K. **(b)** Low-$T$ parts of the $R(T)$ curves show the superconducting transitions for different $N$ (main panel) and the corresponding $T_C(N)$ (inset). Error bars indicate the accuracy of determining $T_C$.

To measure the SC gap, we have fabricated superconductor-insulator-normal metal (SIN) tunnel junctions, where few-layer NbSe$_2$ was mechanically exfoliated from bulk crystals in the same way as for our lateral transport measurements and served as the superconducting (S) electrode. A layer of Au and atomically thin hBN were used as the normal electrode (N) and the insulating barrier (I), respectively. Furthermore, the whole assembly was placed on an hBN substrate, thereby encapsulating NbSe$_2$ between two layers of hBN. Fig. 2a shows the schematics and a typical image of our tunneling devices. Details of the fabrication procedure are explained in Supplementary Note 1. Briefly, few-layer crystals of NbSe$_2$ were transferred onto a PMMA membrane, with a 1-2 layer-thick hBN crystal attached to it beforehand, and then placed onto a thicker (~40 nm) hBN crystal, thereby encapsulating NbSe$_2$ and allowing us to use the top hBN as a tunneling barrier. This provided tunnel junctions with typical areas of ~10 μm$^2$ (Supplementary Fig. S1).

The differential tunneling conductance $G(V_b) = dI(V_b)/dV$ was measured by applying a small AC excitation d$V$ ~50 μV superimposed on a DC bias voltage $V_b$, and detecting the AC current d$I$ between the top (Au) and bottom (NbSe$_2$) electrodes. See Supplementary Fig. S2 for measurement schematics. Typical spectra for different $N$ are shown in Fig.

2b. For all thicknesses $N \geq 2$ we observed spectra typical for SIN tunnel junctions, with the full gap seen as zero conductance for $V_b$ below ~0.5 mV (depending on $N$), and sharp conductance peaks just above the gap. The latter are expected for SIN junctions due to the high density of quasiparticle states at the gap edges[30]. While the presence of a full SC gap is clear for all $N$, its value is markedly reduced for thinner samples, in agreement with their reduced $T_c$. To quantify $\Delta$ for different $N$, we have fitted the measured tunneling conductance $G_{NS}(V_b)$ using the standard expression[30,31]

$$G_{\mathrm{NS}} = dI/dV = \frac{G_{\mathrm{NN}}}{N_{\mathrm{N}}(0)} \int_{-\infty}^{+\infty} N_{\mathrm{S}}(E, \Gamma, \Delta) \frac{\partial f(E+eV_b)}{\partial(eV_b)} dE, \qquad (1)$$

where $G_{\mathrm{NN}}$ is the tunnelling conductance corresponding to both electrodes being in the normal state, $N_{\mathrm{N}}(0)$ and $N_{\mathrm{S}}(E,\Gamma,\Delta)$ is the density of states (DoS) at the Fermi level for the superconducting electrode in the normal and superconducting state, respectively; $f(E + eV_b, T)$ is the Fermi-Dirac distribution, and $\Gamma$ the quasiparticle lifetime broadening. The superconducting DoS is given by the Dynes formula[32] that also applies for ultrathin films[33]

$$N_{\mathrm{S}}(E,\Gamma,\Delta) = Re\left\{\frac{E - i\Gamma}{\sqrt{(E-i\Gamma)^2 - \Delta^2}}\right\}. \qquad (2)$$

where the gap $\Delta$ is assumed to be anisotropic, as extensively discussed in literature[34-38]. See Supplementary Note 2 for further details. The fitting showed that $\Delta \equiv \Delta_{\max}$ ($T$=0.3K) decreases from its bulk value of 1.3 meV for $N$ =50 to $\approx 0.6$ meV for $N$ =2, that is, $\Delta$ exhibited a reduction by more than a factor of 2 – see below for details. Unfortunately and despite many attempts, we were unable to obtain tunneling spectra for monolayer NbSe$_2$. All our monolayer devices showed no detectable tunneling current, probably because of the difficulty to obtain a clean NbSe$_2$-hBN interface for very small available monolayer crystals (exfoliated bilayers were much larger).

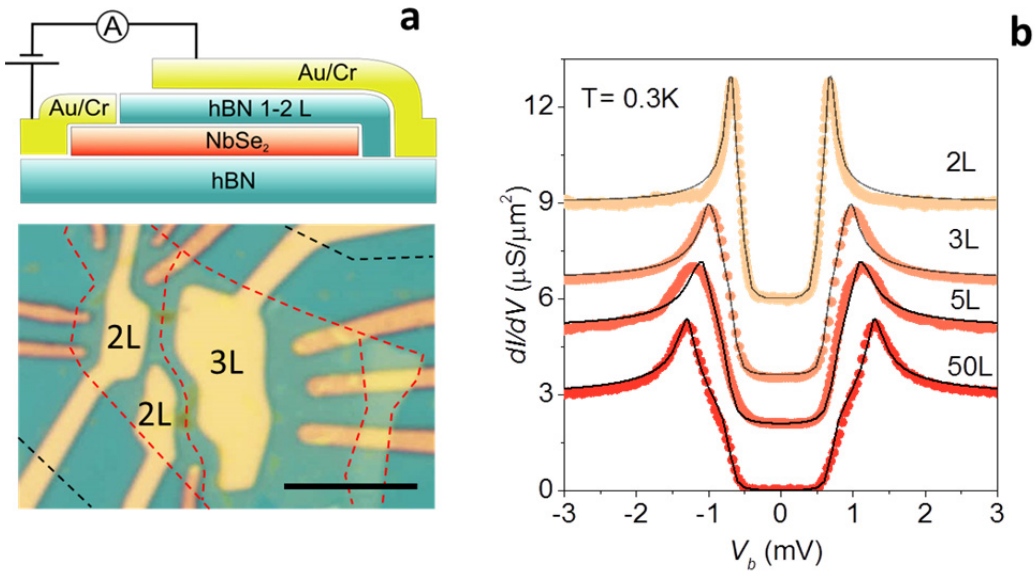

**Figure 2. Superconducting gap in few-layer NbSe$_2$. (a)** Top: sketch of our tunneling devices; bottom: optical image of 2 and 3 layers NbSe$_2$ devices. The red and black dashes outline NbSe$_2$ and thin hBN crystals providing the tunnelling barrier, respectively. Scale bar, 10μm. **(b)** Tunneling conductance d$I$/d$V$ as a function of bias voltage, $V_b$, for crystals with different $N$ (symbols). Black solid curves are fits to eq. (2), see text. The spectra, labelled by $N$, are normalized and shifted for clarity.

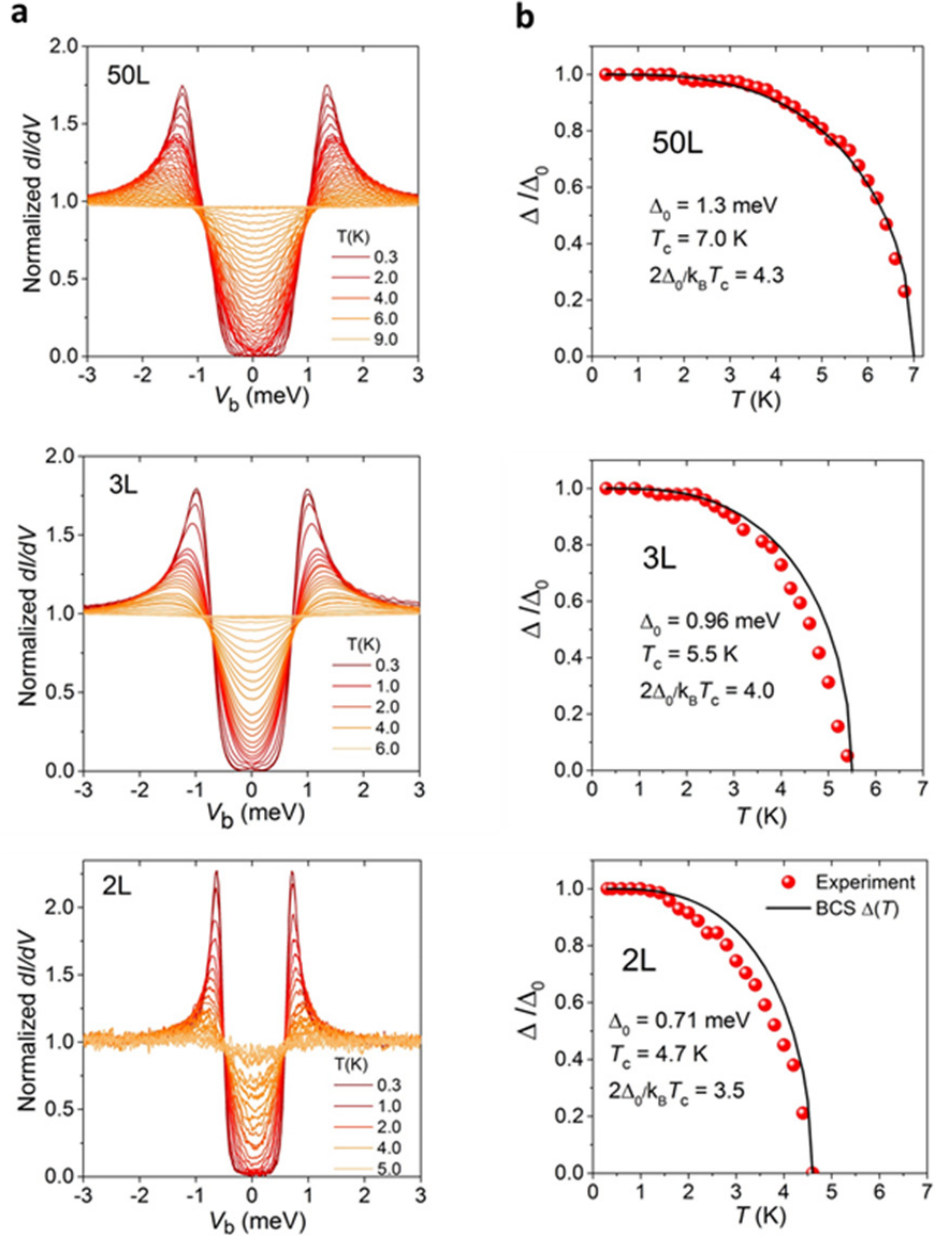

**Figure 3. Temperature dependence of the superconducting gap in few-layer NbSe$_2$. (a)** Typical tunnelling spectra for devices with $N$= 2, 3 and 50 at different temperatures (see labels). **(b)** Symbols show experimental values of $\Delta(T)$ extracted from fitting of the tunnelling spectra in (a). Solid curves are fits to the standard BCS expression $\Delta(T) \propto \tanh(1.74\sqrt{T_c/T - 1})$.

The temperature dependence of the SC gap for different $N$ is shown in Fig. 3. For $N$ =50, $\Delta(T)$ is accurately described by the BCS theory (Fig. 3a). For thinner samples there are some deviations but the agreement is still good (Figs 3b,c). These fits allowed the extrapolation to determine the zero-$T$ gap $\Delta_0$ that for all our samples was found to be indistinguishable (within our accuracy) from $\Delta$ for our lowest measurement temperature of 0.3 K. Importantly, the transition temperatures corresponding to closing of the SC gap agree within $\pm 0.2$ K with $T_C$'s determined from the resistance measurements (cf. Fig. 1b). This further confirms the consistently high quality of few-layer NbSe$_2$ used in our devices.

Before comparing the thickness dependences of the SC transition temperature and the gap in our NbSe$_2$ samples, we discuss the influence of the BKT transition[39,40] on the observed $T_C$, i.e., the effect of thermal fluctuations that become significant in 2D superconductors and lead to the formation of vortex-antivortex pairs even in the absence of an external magnetic field. In this situation vortex unbinding gives rise to dissipation, which in transport measurements is seen as a resistive transition, even though the material is still in the superconducting state. The BKT transition occurs at a critical temperature $T_{\text{BKT}} < T_{c0}$ (here $T_{c0}$ is the intrinsic transition temperature due to Cooper pair breaking). To determine whether the BKT transition plays a role in the observed decrease of $T_c$ for small $N$, we analyzed the temperature-dependent resistance $R(T)$ and $I$-$V$ characteristics of our thinnest crystals, $N$ =1 and 2, in the vicinity of the superconducting transition – see Fig. 4. The standard techniques used to determine $T_{\text{BKT}}$ is either by fitting the low-temperature $R(T)$ curves using e.g. Halperin-Nelson formalism[40,41] that describes the resistivity near $T_{\text{BKT}}$ in the presence of thermally excited free vortices, or by analysing the critical exponents, $\alpha(T)$, describing the voltage–current characteristics, $V \propto I^\alpha$ (ref.[42]). In the former case

$$R(T) \approx 10.8 \cdot c \cdot R \cdot \exp\{-2\sqrt{c(T_{c0} - T_{\text{BKT}})/(T - T_{\text{BKT}})}\}, \quad (3)$$

where $T_{c0}$, $T_{\text{BKT}}$ and dimensionless $c$ are adjustable parameters and $R$ is the measured normal-state sheet resistance. As shown in Fig. 4, for monolayer NbSe$_2$ $T_{\text{BKT}}$ differs from $T_{c0}$ by just 0.2K (≈1.8K vs 2K) while for $N$ = 2 the two temperatures are practically indistinguishable. A similar result was obtained from analysis of the exponent $\alpha(T)$ in $V \propto I^\alpha$. The commonly used criterion[41,42] is $\alpha = 3$ corresponding to $T = T_{BKT}$ and $\alpha = 1$ for $T = T_{c0}$. As seen from Fig. 4, the results are almost identical to the analysis of $R(T)$ dependence, yielding $T_{BKT} \approx T_{c0}$ for both $N = 1$ and 2. We note that such closeness of the two temperatures in our samples is not surprising: The combination of a relatively large mean-free path, $l > \xi_0$, and high carrier concentration in NbSe$_2$, $n \sim 10^{15}$ cm$^{-2}$ per layer (Supplementary Note 3) implies a high superfluid stiffness[40], $K \sim n_s(T) \hbar^2/4mk_BT$, where $n_s(T) \propto n$ is the temperature-dependent superfluid density and $m$ the electron mass. The relative temperature difference $\tau_c = (T_{c0} - T_{BKT})/T_{c0}$ can be estimated as[40] $\tau_c \sim 2k_B T_{c0} \beta m / \pi \alpha' \hbar^2 d$ where $\alpha'$ and $\beta$ are material-dependent coefficients in the GL equations, the ratio $\alpha'/\beta$ corresponds to zero-temperature superfluid density $n_s(0)$ and $d$ is the film thickness. Assuming that in our high-quality crystals the superfluid density is not too different from carrier concentration, that is, $n_s(0) \sim n \approx 10^{22}$ cm$^{-3}$, we obtain for the bilayer $\tau_c \sim 10^{-3}$, in agreement with the observed $T_{c0} \cong T_{BKT}$. For the monolayer, the mean-free path becomes comparable with $\xi_0$ (Supplementary Figure S6b) which should considerably reduce $\alpha'$ (ref.[40]) resulting in a larger $\tau_c$. This explains a somewhat larger $\tau_c \sim 0.1$ for the monolayer (Fig. 4c,d).

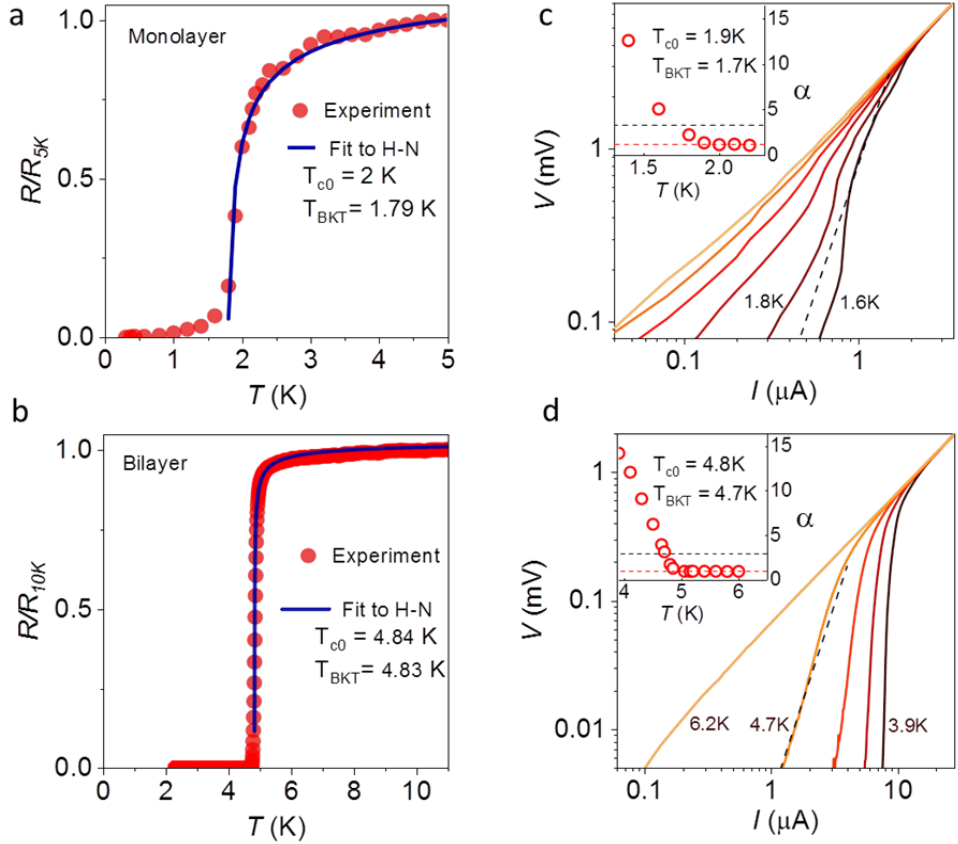

**Figure 4. Role of the BKT transition in mono- and bi- layer NbSe$_2$. (a)** Superconducting transition for monolayer NbSe$_2$ and the corresponding fit to Halperin-Nelson formula (see text). The shown resistance $R$ is normalized to its low-$T$ value in the normal state. **(b)** Same as (a) but for a bilayer NbSe$_2$. **(c, d)** Current-voltage characteristics (shown on a logarithmic scale) for the mono- and bilayer NbSe$_2$ crystals in (a,b) at different $T$. The dashed lines correspond to $V \propto I^\alpha$ with the critical exponent $\alpha$ = 3. The insets show $T$ dependences of $\alpha$ (see text).

The main panel of Fig. 5 compares the thickness dependences of $T_C$ and $\Delta_0$. It is clear that the suppression of $T_C$ follows closely the decrease in the SC gap, indicating that the latter is the main factor determining $T_C$ in our high-quality 2D NbSe$_2$. Note that disorder can in principle reduce $T_C$ significantly as observed in amorphous films of comparable thickness and attributed to weakened Coulomb screening[7,43]. However, this can play only a minor role in our case. Indeed, for this effect to be appreciable, the scattering energy $\hbar/\tau$ should be comparable with the Fermi energy, $E_F$, that is, $E_F \tau/\hbar \approx k_F l \sim 1$ where $\tau$ is the scattering time, $k_F$ the Fermi wavevector and $l$ the mean-free path[43]. This is not the case for any of our devices, where even for the monolayer thickness $k_F l \sim 100$ [using $k_F \approx 0.5$ Å$^{-1}$ (ref. [36]) and $l \approx 15$ nm (supplementary Fig. S6b)]. To quantify the expected reduction in $T_c$ due to this mechanism, we can use the standard theory[43] that describes the disorder effect by two parameters: $t = (e^2/2\pi\hbar)R_n$, the relative value of sheet resistance with respect to the resistance quantum, and $\gamma = 1/\ln(k_B T_{c,bulk}\tau/\hbar)$ that compares the scattering and thermal energies. Using equation (5) of ref.[43] and the known[37] average Fermi velocity for NbSe$_2$, we obtain $T_C/T_{C0} = 0.985$ (here $T_{C0}$ corresponds to bulk NbSe$_2$). This clearly cannot explain our observations of $T_C/T_{C0} \approx 0.3$ for the monolayer, $\approx 0.7$ for the bilayer and so on.

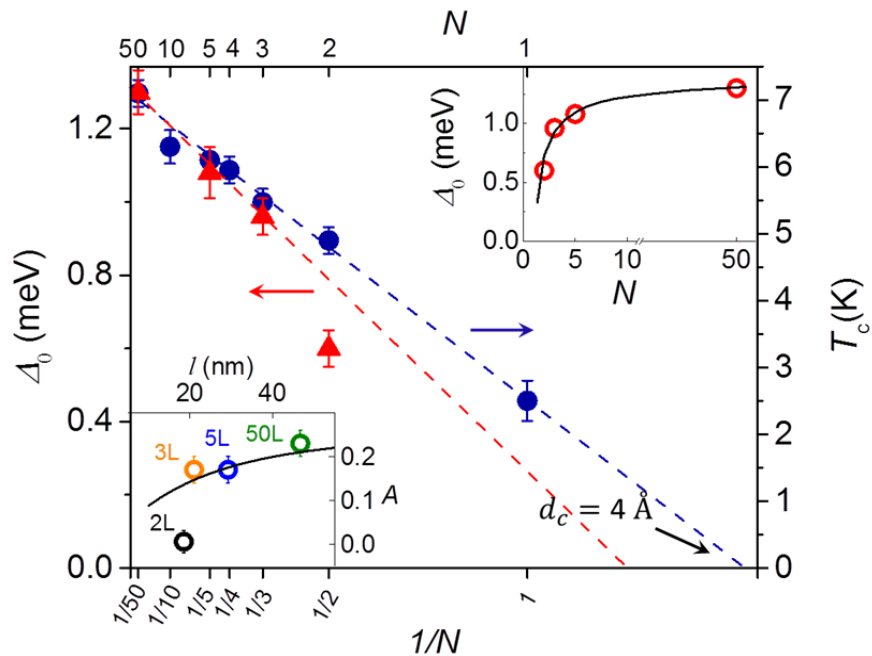

**Figure 5. Thickness dependence of the energy gap and $T_C$ for few-layer NbSe$_2$. Main panel:** Symbols show $\Delta_0 \approx \Delta(0.3K)$ found from tunnelling spectra (red symbols) and the transition temperature $T_C$ found from transport measurements (blue symbols) as functions of $N$. The dashed lines are best fits to $T_C(1/N)$ and $\Delta_0(1/N)$ (for $N = 3, 5$ and 50 only). Error bars for $T_C$ correspond to the width of the SC transition and for $\Delta_0$ to the accuracy of the fit. **Top right inset**: $\Delta_0$ vs $N$. The solid line is a guide to the eye. **Bottom left inset**: Dependence of the anisotropy parameter $A$ on the electron mean-free path, $l$. Symbols: parameter $A$ extracted from fits to the experimental data. Solid curve: $A(l)$ dependence obtained using eq. (S10) from Supplementary Note 4 assuming the bulk anisotropy parameter $A_0 = 0.26$, Fermi velocity $v_F = 1$ eV Å (ref.[35]) and $\Delta_0 = 1.3$ meV. Color-coded labels give the number of layers for each point.

Having ruled out the BKT transition and disorder as possible reasons for the observed suppression of $\Delta$ and $T_C$, we note that, in principle, suppression of the SC gap can also arise from changes in the band structure. For NbSe$_2$, such changes are expected to be rather pronounced in the monolayer limit[44]. In the bulk, NbSe$_2$ has two approximately cylindrical Fermi surfaces corresponding to Nb-derived bands and centered on $\Gamma$ and $K$ points of the Brillouin zone, and a Se-derived pocket around $\Gamma$ point. From analysis of scanning tunneling spectra in ref.[45] it was suggested that an intrinsic SC gap opens only in one of the bands - associated with Nb cylinders around the K point – while the smaller gaps, at $\Gamma$ for Nb and possibly for the Se band, are induced by interband coupling. On the other hand, ARPES experiments[36] suggested that gaps of different magnitude are intrinsic to all Nb bands, with a larger, k-dependent gap at $\Gamma$ and a smaller gap at K points. In all cases the experimentally measured average gap of 1.1 meV was found to be distributed from 0.7 to 1.3-1.4 meV, very similar to the values found for our $N$=50 device (see fits in Supplementary Fig. S4). For $N$ = 1 and 2 the density functional theory[44] suggests significant changes in the Nb bands, with the energy separation between the two bands crossing the Fermi level reduced for the bilayer and one of the bands disappearing altogether for the monolayer. This band-structure reconstruction can also be expected to have a profound effect on the SC gap and, in particular, may be responsible for the noticeable deviation of $\Delta_0(N=2)$ from the overall trend in Fig. 5 (see the discussion below).

For thicker crystals, $N \geq 3$, there are no band structure calculations in the literature but, given the smooth evolution of $\Delta_0$ seen in our experiment for these $N$, it is reasonable to assume that any changes in the band structure for $N > 2$ are relatively small. We believe that they are unlikely to explain the observed suppression of $\Delta_0$ and $T_C$ which is substantial even for $N$ = 10 (Fig. 5). Therefore, we recall another effect that should be general for 2D superconductors but so far was overlooked: In atomically thin superconductors we have to take into account that the wavefunctions of normal electrons that form Cooper pairs have to satisfy the vacuum boundary condition. This results in a gradient of the wavefunction and the corresponding depletion of the superconducting density of states near the surface, as first suggested by de Gennes[46]. This contribution can be described by an additional surface term in the Ginzburg-Landau energy with a boundary condition[46,47]

$$\nabla\Delta \cdot \hat{\mathbf{s}}|_S = -C\Delta_S \qquad (4)$$

where $\Delta(r)$ is the superconducting gap, $\hat{\mathbf{s}}$ the unit vector normal to the surface, $C \approx a/N_0 V \xi^2(0)$, $N_0$ the density of states at the Fermi energy, $V$ the effective pairing potential that includes both attractive electron-phonon interactions and repulsive electron-electron contributions[47,48] and $a$ the Thomas-Fermi screening length, which for metals is of the order of the interatomic distance, $a \approx 1$ Å. The surface energy term and the associated boundary condition (4) can be ignored in bulk superconductors, where it introduces only a small ($1/d$) correction. However, it becomes dominant in 2D crystals, leading to the linear dependence of $T_C$ on the inverse thickness[47]

$$T_C(d) = T_{c,\text{bulk}}(1 - \frac{d_c}{d}) \quad (5)$$

where $d_c = 2C\xi^2(0) \approx 2a/N_0V$ is the critical thickness corresponding to $T_C = 0$. As shown in ref.[47] the constant $C$ effectively compares the film thickness $d$ with the screening length $a$. Once the two become of similar magnitude, the surface suppression of the SC DoS also becomes significant, reducing both the overall value of the gap and $T_C$. Note that in refs.[46,47] this screening effect was discussed for thin films of 3D metals. Although our few-layer NbSe$_2$ is crystallographically 2D, its electronic system remains 3D because of the high electron concentration. Also, the electron wavefunctions for out-of-plane and in-plane momenta may be slightly different due to the layered structure but this should result in little effect on the discussed surface screening.

The predicted $T_C(d) \propto 1/d$ behavior is in good agreement with our observations – see the main panel of Fig. 5. We note that, although both $\Delta(N)$ and $T_C(N)$ are well described by $\propto 1/N \propto 1/d$ dependence, the rate of decrease for $\Delta(N)$ appears to be somewhat greater, with a noticeable deviation for $N$ = 2. This behavior can be attributed to a decrease of the $2\Delta_0/k_B T_C$ ratio with $N$, from 4.3 for $N = 50$ (consistent with its value in the bulk[34]) to ≈3.0 for the bilayer. We speculate that the observed changes in $\Delta_0/T_C$ is another effect, which arises due to the difference in how averaging of the anisotropic SC gap comes through in measurements of the tunneling spectra and the resistive transition. Indeed, in agreement with the previous studies of bulk NbSe$_2$ (e.g., refs.[34-36,45]) the tunneling spectra for our thick samples show strong deviations from the standard BCS behavior: the DoS broadening and kinks in the tunneling spectra correspond to a highly anisotropic gap[49] or, alternatively, to two coupled gaps – a large and a small one – which open at different electronic bands and are coupled by e.g. interband quasiparticle scattering[45]. The gap anisotropy for bulk NbSe$_2$ is also in agreement with expectations from its band structure as known in the literature and described above. However, for our thinner samples, $N$ =3 and 5, the degree of anisotropy noticeably decreases whereas for $N$ = 2 the tunneling spectrum is accurately described by the standard, isotropic BCS expressions (see Supplementary Fig. S5). In principle, one can expect certain washing-out of the anisotropy by enhanced surface scattering in our thinnest devices (Supplementary Fig. S6b)[50]. However, the reduction in the mean-free path $l$ from $N$ = 50 to 2 is smooth and relatively small whereas the disappearance of anisotropy is abrupt and occurs sharply from $N$ = 3 to 2 (Supplementary Fig. S5 and inset of Fig. 5). We speculate that this happens again due to the band structure reconstruction, that is, the same mechanism that is responsible for the stronger than $1/N$ changes in $\Delta_0$ for $N$ = 2. Note that for monolayer NbSe$_2$ the latter mechanism has been corroborated in recent ARPES experiments[19].

Using the known parameters for NbSe$_2$, we estimate the critical thickness $d_c$ in equation (5). The reported value of the electron-phonon coupling constant $\lambda_{e-ph}$ is ≈0.9 (ref.[36]) and the well-known relation $N_0V = \lambda_{e-ph} - \mu^*/1 + \lambda_{e-ph}$ (ref.[48]) yields $N_0V \approx 0.4$ and, therefore, $d_c \approx 5$ Å, where we used a typical value of the Coulomb pseudopotential $\mu^* \approx 0.1$. The estimated $d_c$ matches well the critical thickness obtained experimentally from extrapolation of our $T_C$ vs 1/$N$ data in Fig. 5, $d_c \approx 4$ Å. Interestingly, $d_c$ extrapolated from our data is smaller than the thickness of monolayer NbSe$_2$ of 6.3 Å. This is consistent with the fact that monolayer NbSe$_2$ remains superconducting even in the presence of strong disorder and/or contamination, as indicated by somewhat different $T_c$'s measured by different groups[13,16,19,21].

Finally, we note that the above discussion does not include charge density waves (CDW) that are well known to coexist with superconductivity in NbSe$_2$ (e.g. refs.[19,34-36,51]) and, in principle, may affect $T_C$. To this end, we point out that a recent study of CDWs in few-layer NbSe$_2$ (ref.[21]) found that the CDW transition temperature increased with decreasing number of layers, in contrast to the decreasing $T_C$. No causal link could be established between the opposite changes in the transition temperatures. Furthermore, the view that superconductivity and CDWs should always be in competition is now being challenged,[12,52] with an emerging understanding that the relationship is likely to be more subtle, with incommensurate CDWs possibly favouring superconductivity whereas commensurate CDWs not. For $NbSe_2$, where CDWs are always incommensurate, several reports suggested that CDW and SC are either not in competition[36] or, in fact, CDW may even boost superconductivity[37].

In summary, we measured the thickness dependence of the superconducting energy gap and the transition temperature in 2D $NbSe_2$ crystals. The gap is found to decrease with the number of layers in the same way as the transition temperature. The gap is appreciably anisotropic for our relatively thick crystals ($N > 5$) but for bilayer $NbSe_2$ the anisotropy practically disappears, probably due to the changes in the electronic band structure which were predicted in this case. Based on the observed $\propto 1/N$ dependence, we attribute the pronounced decrease in both $\Delta$ and $T_C$ for small $N$ mostly to the inevitable depletion in the Cooper pair density near the superconductor-vacuum interface (for $N = 2$ there is an additional effect due to changes in the band structure). The absolute value of the surface suppression is in good agreement with the theory. Although this surface effect is insignificant in bulk superconductors, it cannot be neglected in 2D crystals with the thickness comparable with the Thomas-Fermi screening length.

## Supporting information

### 1. Device fabrication.

To fabricate superconductor-insulator-normal metal (SIN) tunnel junctions, bulk single crystals of NbSe$_2$ with typical sizes >10 mm were mechanically exfoliated onto Si/SiO$_x$ substrates coated with polypropylene carbonate (PPC). The thickness of PPC was optimized for high optical contrast, facilitating identification of NbSe$_2$ crystals of desirable thickness (Fig. S1a). In addition, the viscoelasticity of PPC (ref. S1) allowed us to transfer NbSe$_2$ from the polymer directly onto a thin crystal of hexagonal boron nitride (hBN). To this end we used a polymethylmethacrylate (PMMA) membrane with 1-2 layer thick crystals of hBN attached to it previously and the latter was used to 'pick up' NbSe$_2$ from the PPC layer. The thin hBN later served as a high quality tunnel barrier[S2]. The hBN/NbSe$_2$ sandwich was then placed onto a thicker hBN crystal (30-40 nm thick) exfoliated onto another Si/SiO$_x$ substrate, thereby fully encapsulating NbSe$_2$ between two layers of hBN, with the thin layer on the top (Fig. S1b). Each resulting heterostructure was imaged using atomic force microscopy (AFM) in order to select regions of uniform thickness suitable to fabricate tunnel junctions (Fig. S1c). As a final step, a normal-metal electrode (Cr/Au) was patterned using electron beam lithography and thermal evaporation. A second round of lithography was used to define contacts to NbSe$_2$ (Fig. S1d); the thin hBN layer in the contact areas was removed using plasma etching prior to deposition of Cr/Au.

For most lateral transport devices we used the fabrication procedure reported in ref. S3, i.e., NbSe$_2$ was exfoliated directly onto atomically flat SrTiO$_3$ substrates and encapsulated with graphene (this was necessary because NbSe$_2$ cannot be easily lifted from SiO$_2$ with a PMMA/graphene or PMMA/hBN membrane, as is usually done for graphene/hBN heterostructures[S4]). For a given $N$ the device resistance $R$ and its temperature dependence were found to be independent of the encapsulating material (hBN or graphene), presumably because of little charge transfer between graphene and NbSe$_2$. In this situation the significantly higher resistance of undoped graphene (several kΩ vs <200 Ω for NbSe$_2$) ensured that all current flows through NbSe$_2$ whether it is encapsulated by the insulating hBN or highly resistive graphene. As a check, we have prepared a number of devices using the PPC-based transfer method described above. Figure S2 compares the temperature-dependent sheet resistance of two devices where the trilayer NbSe$_2$ ($N$ =3) is encapsulated on both sides with hBN (red curve) and is deposited onto SrTiO$_3$ and encapsulated with graphene (green). It is clear that the superconducting transition is virtually identical in both cases.

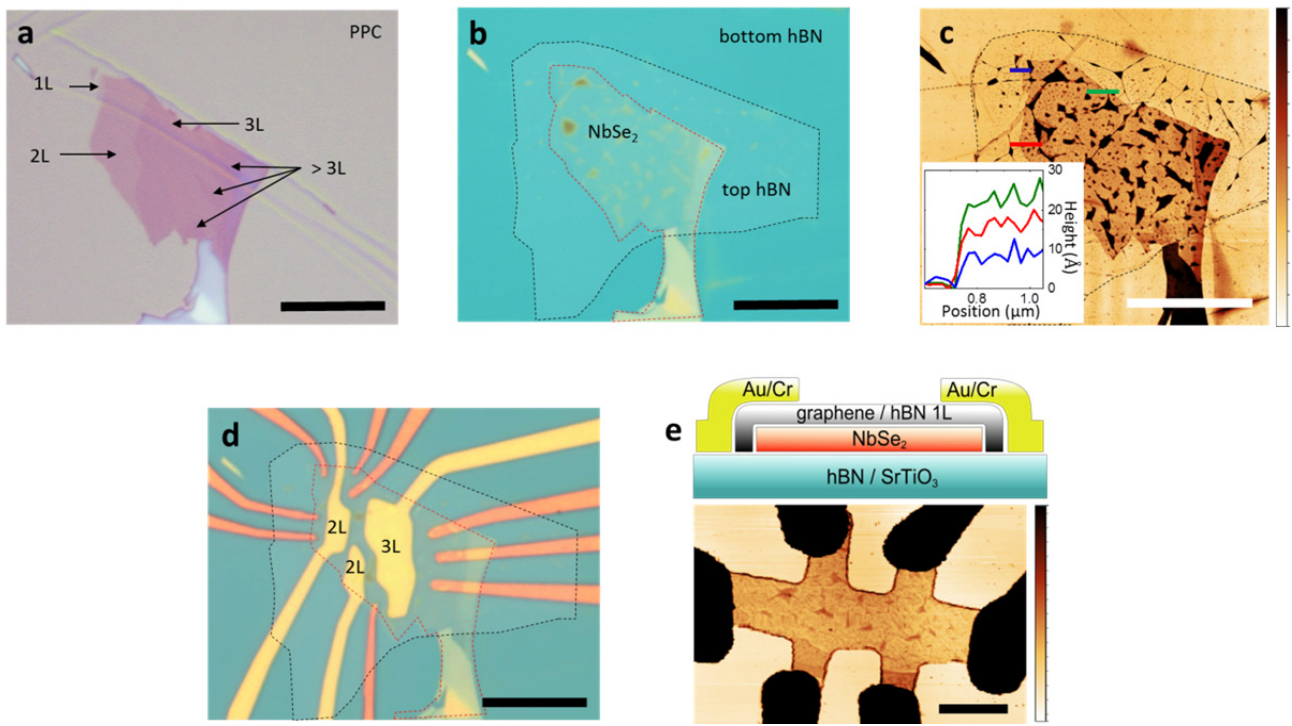

***Figure S1. Device fabrication.*** *(a) Optical image of a typical NbSe$_2$ flake exfoliated onto a PPC/SiO$_2$/Si substrate (the shown flake was used to make tunneling devices with N=2 and 3). Arrows indicate regions of different thickness. The very small size of the monolayer seen on the left is quite typical and is the most likely reason for our failure to make a N=1 device with detectable tunnelling current: as can be seen by comparing with (c), the monolayer part is smaller than typical contamination-filled 'bubbles', making it virtually impossible to obtain a clean NbSe$_2$-hBN interface, as it was always partially or even fully covered by a bubble. **(b)** NbSe$_2$ flake from (a) (highlighted in red) encapsulated with thin hBN (highlighted in black) and transferred onto an hBN substrate. **(c)** AFM image of the same flake; the minimum value of the color scale corresponds to 0 and the maximum to 20 nm. The inset shows height profiles of NbSe$_2$ corresponding to areas with 1, 2 and 3 layers. **(d)** Finished tunnel junction. Orange color highlights the contacts to NbSe$_2$, and yellow shows the top Au electrodes. **(e)** Sketch of a lateral transport device, encapsulated with graphene or 1L hBN (top); AFM image of a monolayer NbSe$_2$ device used in our transport measurements. The monolayer was not attached to thicker flakes. Scale bars in (a)-(d) correspond to 10 µm and in (e) to 1µm.*

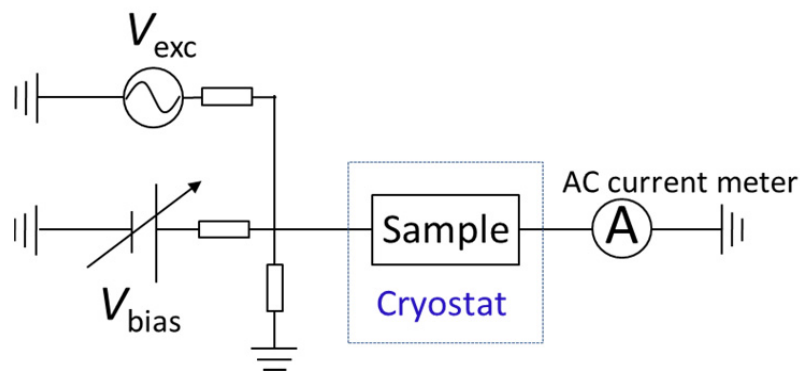

***Figure S2. Measurement schematics and details.*** *Shown is the circuit diagram used for differential conductance measurements. Measurements were carried out in a $^3$He cryostat with variable temperature between 0.3 K to100 K. Lock-in amplifier (SR830, Stanford research system) was used to supply the AC excitation and measure the AC current. Keithley 2614B source-meter was used as the DC source. AC voltage (0.1 V, 16.666 Hz) was applied to the 1: 2000 divider with 5 Ω impedance, which provided 50 mV AC excitation applied to the sample.*

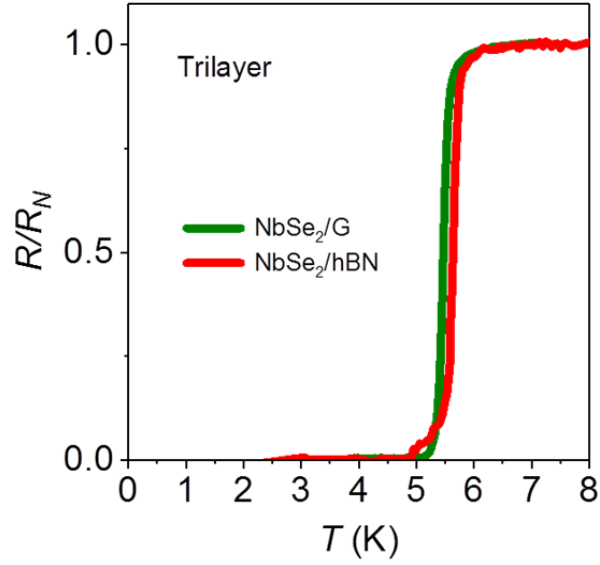

*Figure S3. Comparison of the resistive superconducting transition for NbSe$_2$ encapsulated with graphene and hBN (see text).*

## 2. Fitting of the tunneling spectra.

The differential conductance of a normal metal-insulator-superconductor (NIS) junction can be described as[S5]:

$$G_{NS} = dI/dV = \frac{G_{NN}}{N_N(0)} \int_{-\infty}^{+\infty} N_S(E, \Gamma, \Delta) \frac{df(E+eVb, T)}{d(eVb)} dE \quad (S1)$$

Here $G_{NN}$ is the tunnelling conductance corresponding to both electrodes being in the normal state, $N_N(0)$ and $N_S(E, \Gamma, \Delta)$ are the density of states (DoS) at the Fermi level of the superconducting electrode in the normal and superconducting state, respectively, and $f(E + Vb, T)$ the Fermi-Dirac distribution. The superconducting density of states is given by the Dynes formula[S6]:

$$N_S(E, \Gamma, \Delta) = Re\left\{\frac{E - i\Gamma}{\sqrt{(E - i\Gamma)^2 - \Delta^2}}\right\} \quad (S2)$$

where $\Gamma$ is the quasiparticle lifetime broadening and $\Delta$ the superconducting energy gap.

In the case of an isotropic, multi-gap superconductor (e.g. different gaps around $\Gamma$ and $K$ points in the Brillouin zone), the total DoS is given by a weighted sum:

$$N_S(E, \Gamma_1, \Delta_1, \Gamma_2, \Delta_2) = Re\left\{C \frac{E - i\Gamma_1}{\sqrt{(E - i\Gamma_1)^2 - \Delta_1^2}} + (1 - C) \frac{E - i\Gamma_2}{\sqrt{(E - i\Gamma_2)^2 - \Delta_2^2}}\right\} \quad (S3)$$

where $\Delta_1, \Delta_2$ are the two gaps, $\Gamma_1, \Gamma_2$ the corresponding quasiparticle lifetime broadening, and $C$ the contribution from different pockets of the Fermi surface.

For a superconductor with a single anisotropic gap characterized by an anisotropy parameter $A$, the $k$ dependence of $\Delta$ can be expressed through the polar angle $\theta$ (ref. S7). As the Fermi surface of NbSe$_2$ around the $\Gamma$ point has six-fold symmetry[S9-S11] the angular dependence of $\Delta$ can be written as:

$$\Delta(\theta, A) = \Delta_0[A \cdot cos(6\theta) + (1 - A)] \quad (S4)$$

where $\Delta_0$ is the maximum gap value. The average DoS is then

$$N_S(E, \Gamma, \Delta, A) = Re\{\frac{1}{2\pi}\int_0^{2\pi} \frac{E - i\Gamma}{\sqrt{(E - i\Gamma)^2 - \Delta(\theta, A)^2}} d\theta\} \quad (S5)$$

In literature angle-resolved photoemission and scanning tunneling spectroscopy data for $NbSe_2$ have been interpreted as indications of either two isotropic gaps[S8] or one or more anisotropic gaps[S9-S11]. Therefore we compared fitting of our tunneling spectra using the corresponding models for the DoS, equations (S2), (S3) and (S5) respectively. The results are compared in Figure S3 for a bulk-like $NbSe_2$ sample ($N = 50$). It is clear that the experimental d$I$/d$V$ curves cannot be accurately described by the standard BCS expression for a single isotropic gap (Fig. S3a). An assumption of two slightly different isotropic gaps (Fig. S3b) provides a better fit, but still not as good as a single anisotropic gap (Fig. S3c). Importantly, the latter accurately reproduces not only the overall shape of the measured tunneling spectrum, but also the characteristic kinks typically seen for bulk $NbSe_2$ (refs. S8,S9). The kinks appear as double peaks in the derivative of d$I$/d$V$ as shown in the inset of Fig. S3c. To further demonstrate that the theoretical description with a single anisotropic gap provides the best fit to all our experimental spectra, Fig. S4 compares the fits with single isotropic and single anisotropic gap for devices with $N$ =2, 3, 5 and 50. For $N$ = 3, 5 and 50 the assumption of an anisotropic gap clearly provides the best fits, with the anisotropy parameter $A$ decreasing for thinner crystals. At the same time, for $N$ = 2 the anisotropy reduces to almost zero, that is, the tunnelling spectrum is described by the standard, isotropic BCS expressions (see main text and Supplementary Note 4 for a discussion).

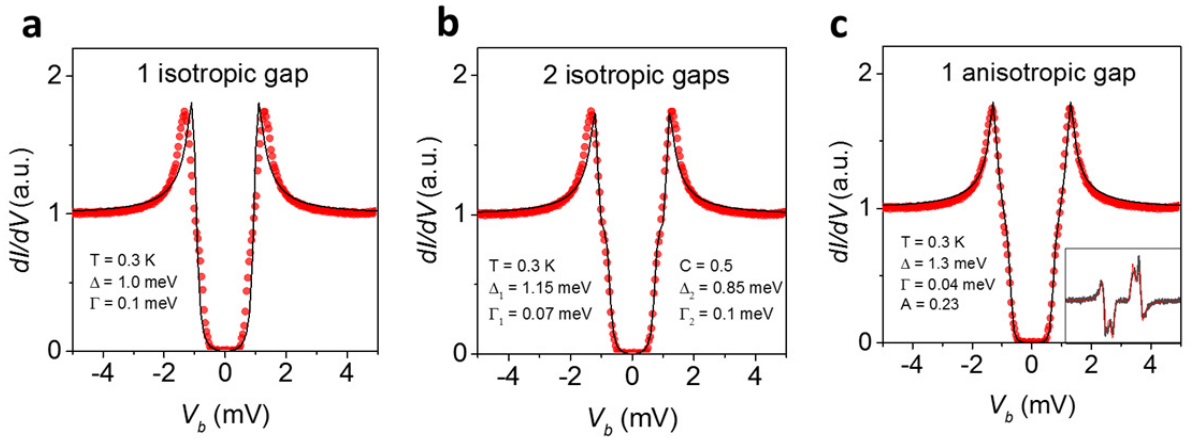

*Figure S4. Best fits to the measured tunnelling spectra for bulk $NbSe_2$ (N=50) with different assumptions. Symbols are experimental data and black solid lines are fits to eq. (S1) using the DOS corresponding to a single isotropic gap (a), two isotropic gaps with equal weights (b) and one anisotropic gap (c). The inset in (c) shows the derivative of the experimental spectrum (black line) and the corresponding fitting curve (red). All shown experimental spectra were obtained at temperature T=0.3 K.*

| NbSe$_2$ thickness | SC gap $\Delta_0$ (meV) | Anisotropy $A$ | QP lifetime broadening $\Gamma$ (meV) | $T_C$ (K) | $2\Delta_0/k_B T_C$ |
|---|---|---|---|---|---|
| **50L** | 1.3 | 0.23 | 0.04 | 7.0 | 4.31 |
| **5L** | 1.1 | 0.17 | 0.07 | 6.2 | 4.12 |
| **3L** | 0.96 | 0.17 | 0.04 | 5.5 | 4.05 |
| **2L** | 0.6 | 0.006 | 0.02 | 4.8 | 2.9 |

*Table S1. Fitting parameters corresponding to the description of experimental tunnelling spectra using the anisotropic single gap model.*

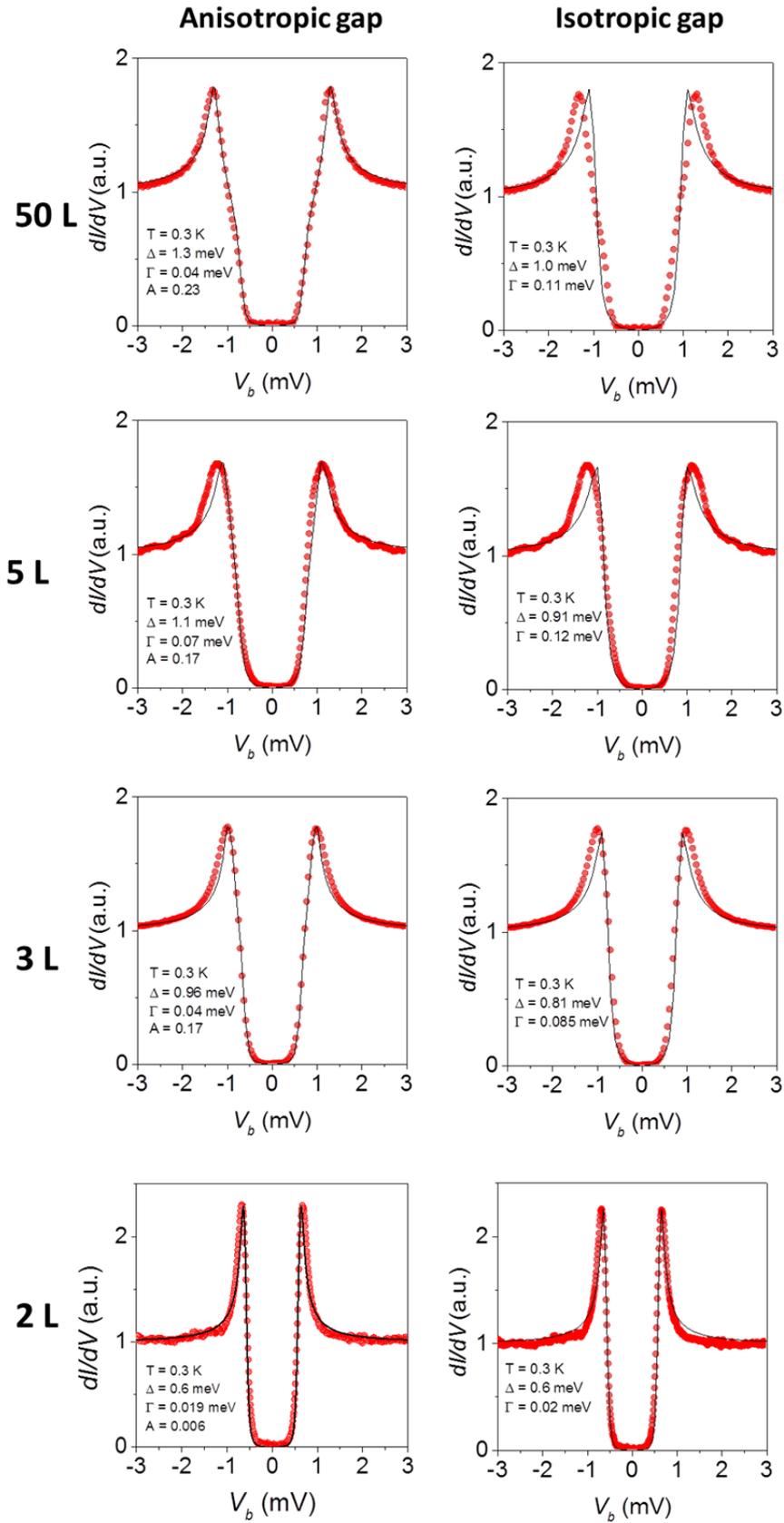

*Figure S5. Fitting of the experimental tunneling spectra with and without taking into account gap anisotropy.* All experimental curves correspond to $T = 0.3$ K. Fitting parameters are shown as legends.

# 3. Estimation of the mean free path for samples with different N

To extract the carrier concentration in our few-layer NbSe$_2$ crystals, we carried out Hall measurements on several samples encapsulated with insulating hBN. As an example Fig. S5a shows the measured Hall resistance $R_{xy}$ vs perpendicular magnetic field $B$ for trilayer NbSe$_2$. This yielded carrier concentration per layer $n \approx 8.8 \cdot 10^{14}$ cm$^{-2}$, in excellent agreement with literature values for bulk NbSe$_2$, $n \approx 8.5\ 10^{14} cm^{-2}$ (ref. S12). Here we used measurement temperatures $T > 40$ K, as the Hall coefficient ($R_H$) of NbSe$_2$ is known to be constant in this temperature range (ref. S13) and therefore provides a good estimate of the total carrier density[S12]. To estimate the mean-free path $l$, we have used the fact that lateral transport in bulk NbSe$_2$ is known to have quasi-2D character, due to the open cylindrical structure of the Fermi surface. This allows considering atomically thin crystals of NbSe$_2$ as stacks of individual 2D conductors and estimating $l$ within individual layers using the 2D Drude formula:

$$l_{2D} = \frac{h}{e^2} \frac{\sigma_{2D}}{\sqrt{2\pi n_0}} \quad (S6)$$

Here the conductance $\sigma_{2D} = 1/NR$ is taken equal to the inverse low-$T$ sheet resistance per layer and $n_0$ is the carrier concentration per layer. The calculated dependence of $l$ on the number of layers $N$ is shown in Fig. S5b. The mean-free path is reduced for the thinner crystals due to additional scattering at the surfaces (see main text), but for all $N$ it remains larger that the SC coherence length reported in literature, $\xi_0 \sim 8$ nm (ref. S14).

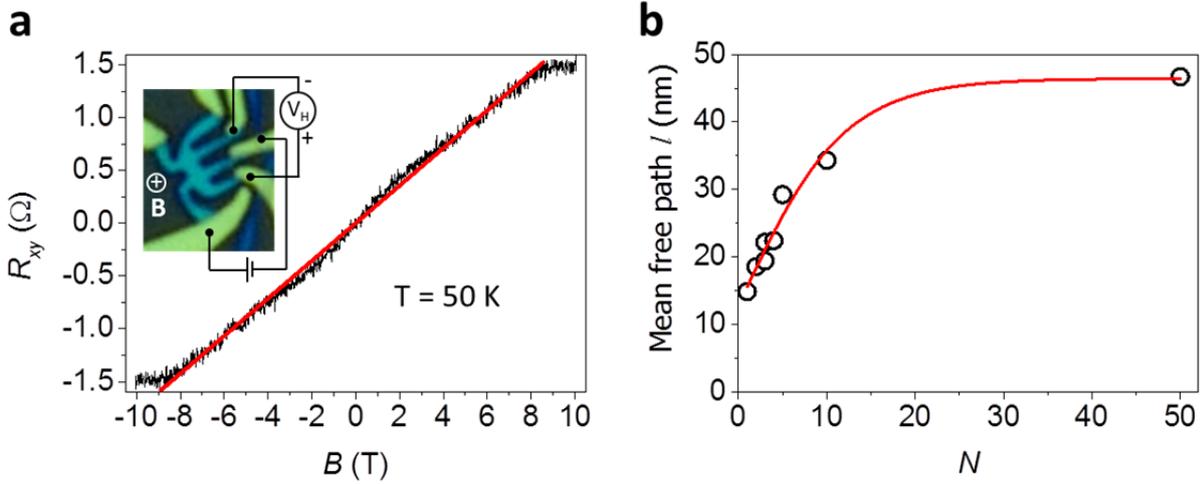

*Figure S6. Mean free path for NbSe$_2$ of different thickness. (a) Hall resistance $R_{xy}$ as a function of magnetic field, B (black); red line is linear fit yielding $R_H$ = 0.178 m$^2$/C. The inset shows an optical microscope image of the device and schematics of the Hall measurements. (b) Mean free path $l$ as a function of N calculated using (S6). Solid line is a guide to the eye.*

# 4. Anisotropy of the superconducting gap and its relation to the mean-free path.

To better understand the observed smearing of the gap anisotropy in our thinner crystals, we have analyzed the expected relation between the decrease of the anisotropy parameter $A$ and the electron mean-free path in our NbSe$_2$ crystals with different $N$. As described in the previous section, the mean-free path gradually decreases from $l \approx 50$ nm for *N* =50 to $l \approx 15$ nm for the monolayer. On the other hand, scattering by non-magnetic defects is known to decrease the gap anisotropy, especially in SC with the mean-free path larger than the coherence length[S15], as is the case in our NbSe$_2$ crystals.

In superconductors with finite gap anisotropy, the angular distribution of gap values, $\Delta(\theta)$, can be related to the average gap accessible experimentally via the normalized mean-square gap anisotropy $\langle a^2 \rangle$ (ref. S16):

$$\langle a^2 \rangle = \frac{\langle (\Delta(\theta) - \overline{\Delta})^2 \rangle}{\overline{\Delta}^2} \qquad (S7)$$

where $\langle \ \rangle$ denotes the angular average over the gap distribution and $\overline{\Delta}$ the average SC gap. Using eq. (S4) for the six-fold symmetric gap of NbSe$_2$ this becomes:

$$\langle a^2 \rangle = \frac{A^2}{2(1-A)^2} \qquad (S8)$$

where *A* is the anisotropy parameter. In the presence of additional surface scattering that reduces the mean free path $l$ in few-layer crystals of NbSe$_2$ (see main text for a discussion), the mean-square gap anisotropy changes from its bulk value $\langle a^2 \rangle_0$ to approximately[S16]:

$$\langle a^2 \rangle \approx \frac{\langle a^2 \rangle_0}{[1+\left(\hbar v_F / 2l\overline{\Delta}\right)^2]} \qquad (S9)$$

Combining (S8) and (S9) we obtain the following equation for the anisotropy parameter *A* as a function of the mean-free path $l$:

$$\frac{A^2}{2(1-A)^2} = \frac{\langle a^2 \rangle_0}{[1+\left(\hbar v_F / 2l\Delta_0(1-A)\right)^2]} \qquad (S10)$$

The corresponding relation between A and $l$ is shown in the inset of Figure 5 in the main text together with the values of *A* extracted from the measured tunnelling spectra. A clear correlation between the two for *N* = 3, 5 and 50 indicates that the reduced anisotropy can indeed be explained by the increased surface scattering, while for *N* = 2 a more important factor must be the reconstruction of the band structure (see main text).